\documentclass[a4paper]{jpconf}
\usepackage{graphicx}
\usepackage{amssymb}
\usepackage{amsmath, amscd}
\newcommand{\R}{\mathbb{R}}

\begin{document}

\title{Can the Schr{\"o}dinger dynamics explain measurement?}

\author{Alexey A. Kryukov}

\address{Department of Mathematics \& Natural Sciences, University of Wisconsin-Milwaukee, USA}

\ead{kryukov@uwm.edu}

\begin{abstract}
The motion of a ball through an appropriate lattice of round obstacles models the behavior of a Brownian particle and can be used to describe measurement on a macro system. On another hand, such motion is chaotic and a known conjecture asserts that the Hamiltonian of the corresponding quantum system must follow the random matrix statistics of an appropriate ensemble. We use the Hamiltonian represented by a random matrix in the Gaussian unitary ensemble to study the Schr{\"o}dinger evolution of non-stationary states. For Gaussian states representing a classical system, the Brownian motion that describes the behavior of the system under measurement is obtained. For general quantum states, the Born rule for the probability of transition between states is derived. It is then shown that the Schr{\"o}dinger evolution with such a Hamiltonian models measurement on macroscopic and microscopic systems, provides an explanation for the classical behavior of macroscopic bodies and for irreversibility of a measurement, and identifies the boundary between micro and macro worlds.
%
%
\end{abstract}







Behavior of macroscopic bodies under a measurement can be described by Newtonian dynamics, at least in principle. For instance, suppose the position of a point-like particle is measured. The distribution of the position random variable can be 
explained by interaction of the particle with the surroundings and the measuring device. In principle, the effect of this interaction on the particle can be described by the Newtonian equations of motion. Because of the very large number of fluctuations in the position of the particle during the period of observation, it is more realistic to describe the measurement stochastically. 
For instance, the Langevin equation 
can be derived from the Newtonian dynamics of the particle in a medium  and is well-suited for modeling a measurement of the position of the particle. It predicts the normal distribution of the position random variable during the period of observation, which is consistent with experience.


The situation in quantum theory is arguably very different. The common wisdom is that the Schr{\"o}dinger equation is not sufficient to describe measurement in the micro world. The decoherence theory is seemingly the closest one can get to describing a measurement while using the Schr{\"o}dinger dynamics.
However, the decoherence theory does not explain how a particular measured state of the system comes to life and does not provide by itself the probability of an outcome. A modification of the Schr{\"o}dinger equation or an altogether different mechanism seems to be needed to model the process of measurement.
The difficulty lies in the linear property of the Schr{\"o}dinger equation. Even if under the Schr{\"o}dinger evolution a given state were to converge to an eigenstate of the measured observable, the sum of two such states would not do the same. This issue does not arise in Newtonian physics, which is not linear, pointing to its apparent advantage in describing measurement. 

To better understand the difference between two theories in their application to measurement, let us express the relationship between Newtonian and Schr{\"o}dinger dynamics in precise terms.  Let $M^{\sigma}_{3}$  be the set of all Gaussian states
\begin{equation}
\label{g}
g_{{\bf a}, \sigma}({\bf x})=\left(\frac{1}{2\pi\sigma^{2}}\right)^{3/4}e^{-\tfrac{({\bf x}-{\bf a})^{2}}{4\sigma^{2}}}
\end{equation} 
and let $M^{\sigma}_{3,3}$ be the set of all Gaussian wave packets
\begin{equation}
\label{phi}
\varphi_{{\bf a}, {\bf p}, \sigma}({\bf x})=\left(\frac{1}{2\pi\sigma^{2}}\right)^{3/4}e^{-\tfrac{({\bf x}-{\bf a})^{2}}{4\sigma^{2}}}e^{i{\bf p}{\bf x}/\hbar}.
\end{equation}
Here, ${\bf x}, {\bf a}$ and ${\bf p}$ are vectors in $\R^3$ and the variance $\sigma^2$ is assumed to be as small as needed. In quantum mechanics, the wave packets (\ref{g}) and (\ref{phi}) can be used to represent particles of position ${\bf a}$ and velocity ${\bf p}/m$.

The sets $M^{\sigma}_{3}$ and $M^{\sigma}_{3,3}$ can be considered submanifolds of the projective space of states  $CP^{L_{2}}$ of the particle with the induced manifold structure. Moreover, the Fubini-Study metric on $CP^{L_{2}}$ gives rise to a Riemannian metric on  $M^{\sigma}_{3}$ and $M^{\sigma}_{3,3}$. The map $\omega: {\bf a} \longrightarrow g_{{\bf a}, \sigma}$ identifies the classical space $\R^3$ with the submanifold  $M^{\sigma}_{3}$. Likewise, the map $\Omega: ({\bf a}, {\bf p}) \longrightarrow g_{{\bf a}, \sigma}e^{i{\bf p}{\bf x}/\hbar}$ identifies the classical phase space $\R^3 \times \R^3$ with  $M^{\sigma}_{3,3}$. A simple calculation demonstrates that the induced Riemannian metric on  $M^{\sigma}_{3}$ and  $M^{\sigma}_{3,3}$ is the usual Euclidean metric, so that $\omega$ and $\Omega$ are isometric embeddings of the Euclidean classical space and the Euclidean phase space into $CP^{L_{2}}$.

We now claim that Newtonian dynamics is the Schr{\"o}dinger dynamics of the particle whose state is constrained to the manifold $M^{\sigma}_{3,3}$. In that sense, it is similar to a classical dynamical system with a constraint, e.g., bead on a wire. In fact, the action functional
\begin{equation}
\label{SS}
S[\varphi]=\int \overline{\varphi}({\bf x},t) \left[i\hbar \frac{\partial}{\partial t}-{\widehat h}\right] \varphi({\bf x},t) d^3 {\bf x} dt
\end{equation}
with the Hamiltonian ${\widehat h}=-\frac{\hbar^{2}}{2m}\Delta+V({\bf x},t)$ yields the Schr{\"o}dinger equation. For the states $\varphi$ constrained to the manifold $M^{\sigma}_{3,3}$, this functional reduces to the classical action
\begin{equation}
S=\int \left[{\bf p}\frac{d {\bf a}}{dt}-h({\bf p},{\bf a},t)\right]dt,
\end{equation}
where $h({\bf p},{\bf a},t)=\frac{{\bf p}^2}{2m}+V({\bf a},t)+const$ is the Hamiltonian function for the particle. We used here the fact that the parameter $\sigma$ can be made arbitrarily small and that the terms $g^2_{{\bf a}, \sigma}$ form a delta sequence as $\sigma$ approaches $0$.
It follows that the variation of the functional (\ref{SS}) subject to the constraint that $\varphi$ belongs to $M^{\sigma}_{3,3}$ yields Newtonian equations of motion for the particle.

Because Gaussian states form a complete set in the Hilbert space $L_2(\R^3)$, one can prove the converse statement. Namely, there is a unique unitary evolution of state of the particle in $L_2(\R^3)$ that reduces to Newtonian motion when constrained to $M^{\sigma}_{3,3}$. This is exactly the Schr{\"o}dinger evolution with the Hamiltonian ${\widehat h}=-\frac{\hbar^{2}}{2m}\Delta+V({\bf x},t)$ \cite{KryukovPosition}. Furthermore, both statements can be generalized to include systems of $N$ particles interacting via potential $V({\bf x}_1, ..., {\bf x}_N, t)$ \cite{KryukovMacro}.
Thus, we have an isometric embedding of the phase space of an arbitrary classical mechanical system into the space of quantum states of the system 
with the property that the Schr{\"o}dinger dynamics on the space of states is a unique unitary extension of the Newtonian dynamics on the phase space.

The obtained relationship between Newtonian and Schr{\"o}dinger dynamics can be complemented by a relationship between the normal probability distribution for the position of a particle in $\R^3$ and the Born rule for the probability of transition between states in the space of states $CP^{L_{2}}$. 
The existence of such a relationship is clear already from the fact that for the states $g_{{\bf a}, \sigma}$ in $M^{\sigma}_{3}$ the probability density $|g_{{\bf a}, \sigma}|^2$ in the Born rule is also the normal probability density function on $\R^3$. So, the Born rule on $CP^{L_{2}}$ implies the normal probability distribution on $\R^3=M^{\sigma}_{3}$ and is identical to it on this set.

To see this in more detail and to derive the converse statement, 
let us express the Born rule in terms of the probability of transition between normalized states.  Let $\theta(g_{{\bf a}, \sigma}, g_{{\bf b}, \delta})$ be the Fubini-Study distance between the Gaussian states $g_{{\bf a}, \sigma}$ and  $g_{{\bf b}, \delta}$. Let $({\bf a}-{\bf b})^{2}$ be the square of the Euclidean distance between the corresponding points ${\bf a}$ and ${\bf b}$ in $\R^3$. By a direct integration, we have:
\begin{equation}
\label{del}
\left(\frac{2\sigma \delta}{\sigma^2+\delta^2}\right)^3 e^{-\frac{({\bf a}-{\bf b})^2}{2(\sigma^2+\delta^2)}}=\cos^{2}\theta(g_{{\bf a}, \sigma}, g_{{\bf b}, \delta}).
\end{equation}
If $\delta=\sigma$, this equation yields a relationship of the distances between  $g_{{\bf a}, \sigma}$ and  $g_{{\bf b}, \sigma}$ in the Fubini-Study metric on $CP^{L_{2}}$ and between the points ${\bf a}$ and ${\bf b}$ in the Euclidean metric on $\R^3$:
\begin{equation}
\label{mainO}
e^{-\frac{({\bf a}-{\bf b})^{2}}{4\sigma^{2}}}=\cos^{2}\theta(g_{{\bf a}, \sigma}, g_{{\bf b}, \sigma}).
\end{equation}
On another hand, when $\delta$ approaches $0$, the left side of (\ref{del}) yields the normal probability density function times the volume element $(8\pi)^\frac{3}{2}\delta^3$. 
%
%
%
%
%
%
The resulting probability can be interpreted as the probability of finding the particle near point ${\bf b}$, given its initial position at ${\bf a}$. The probability on the right side is the probability of transition between the corresponding initial and end-states, calculated by the Born rule. Once again, we see that the normal probability distribution and the Born rule are identical on $M^{\sigma}_{3}$. In particular, assuming the normal probability distribution of the position on $\R^3$, we deduce the validity of the Born rule on $M^{\sigma}_{3}$. 

Suppose the probability of transition between states depends only on the Fubini-Study distance between them. Then, the validity of the Born rule for the states in  $M^{\sigma}_{3}$ signifies its validity for arbitrary quantum states. In fact, the Fubini-Study distance between states $g_{{\bf a}, \sigma}$ and  $g_{{\bf b}, \delta}$ with ${\bf a}$ and ${\bf b}$ in $\R^3$ takes on all possible values  in $CP^{L_{2}}$ from $0$ to $\pi/2$. Let then $\varphi$ and $\psi$ be any two states in $CP^{L_{2}}$ and 
let $g_{{\bf a}, \sigma}$ and $g_{{\bf b}, \delta}$ be two states at the same Fubini-Study distance as the distance between $\varphi$ and $\psi$.
 The probability of transition between $\varphi$ and $\psi$ is then given by
\begin{equation}
\label{dist}
P_{\varphi, \psi}=P_{g_{{\bf a}, \sigma}, g_{{\bf b}, \delta}}=\cos^{2}\theta(g_{{\bf a}, \sigma}, g_{{\bf b}, \delta})= 
  \cos^{2}\theta(\varphi, \psi).
\end{equation}
So, $P_{\varphi, \psi}=\cos^{2}\theta(\varphi, \psi)$, which is the Born rule. We conclude that under these conditions, the normal probability distribution on $\R^3$ implies the Born rule on the space of states.



The derived relationship between Newtonian and Schr{\"o}dinger dynamics and between normal probability distribution and the Born rule indicates that there may exist a fundamental connection between measurements in the macro and micro worlds. As stated earlier, the process of measurement in classical physics does not require a separate mechanism and can be modeled within Newtonian dynamics itself. In particular, the Langevin equation for the Brownian particle can be derived from the Newtonian dynamics of the particle interacting with a harmonic oscillator heat bath  \cite{Zwanzig}. Because Newtonian dynamics is a constrained Schr{\"o}dinger dynamics, 
%
%
%
it is reasonable to look for a Hamiltonian that yields a Newtonian model of classical measurement of the particle under the constraint. 
%
%
Assuming that such a Hamiltonian exists, we shall 
use it to model measurements in the micro and macro worlds and study the outcomes.


%

To find a proper Hamiltonian, let us model the behavior of a macroscopic particle under measurement by a Brownian motion on $\R^3$.
%
%
Under the embedding $\omega: \R^3 \longrightarrow CP^{L_{2}}$, the motion of the particle is the motion of its state, constrained to the submanifold $M^{\sigma}_{3}$. 
%
%
We therefore need a Hamiltonian that accounts for the  effect of the surroundings and the device on the particle's state and that models the Brownian motion when constrained to $M^{\sigma}_{3}$. 
%
%

Note that the physical Brownian motion can be identified with a regular, stochastic or a chaotic process \cite{Cecconi}.
%
Based on the rich experimental data and the works of Wigner \cite{Wigner} and BGS \cite{BGS}, the general consensus currently is that Hamiltonians of all generic quantum systems whose underlying classical dynamics is chaotic follow the random matrix statistics of an appropriate ensemble. This includes complex systems such as the heavy nuclei considered in \cite{Wigner} as well as simple one-particle systems such as the electron in a lattice of round obstacles. The latter case is particularly relevant to the issue of measurement as the classical motion of a free particle through an appropriate lattice models the Brownian motion of the particle and yields the normal probability distribution for the position  \cite{Cecconi}. 
%

%

%
%

This independent argument suggests that the Hamiltonian that we are looking for exists and can be represented at any time by a random matrix.
However, instead of the usual study of distribution of eigenvalues of a random matrix, our goal here is to investigate the evolution of non-stationary states of a measured particle. Under the evolution with such a Hamiltonian, the state becomes a random variable performing a random walk on the space of states $CP^{L_{2}}$. 
%

To ensure that the Hamiltonian is Hermitian, the random matrix will be assumed to take values in the Gaussian unitary ensemble. The Hamiltonian also needs to account for the independence of action of the surroundings on the particle at different and sufficiently distant moments of time. To summarize, we conjecture that:
%
%
\begin{quote}{\bf{(RM)}}
{\it The Hamiltonian of a quantum system whose underlying classical dynamics describes a Brownian motion
can be represented at any time by a random matrix from the Gaussian unitary ensemble. Matrices representing the Hamiltonian at two different moments of time are independent.}
 %
\end{quote}
This conjecture is essentially the BGS-conjecture \cite{BGS} applied to a specific classical chaotic system, such as a particle in a lattice of round obstacles. The Brownian motion in the conjecture is appropriate for representing the process of measurement in the macro world. This, together with abundant experimental confirmation of the BGS-conjecture suggest that {\bf (RM)} may describe accurately what in fact is happening in a measurement. However, for our purposes it will be sufficient to consider {\bf (RM)} as a way to specify the model of measurement studied in the paper.

Let us prove, first of all, that we found a proper Hamiltonian. Namely, let us show
that the motion of state driven by the Hamiltonian ${\widehat h}$ in {\bf (RM)} and conditioned to stay on $M^{\sigma}_{3}$ yields a random walk on $\R^3$ that approximates the Brownian motion and models the process of  measurement on a macroscopic particle.
In fact, for small time intervals $\Delta t=t_{k}-t_{k-1}$, the state $\varphi_{t_{N}}$ at time $t_N$ is approximately given by the time ordered product
\begin{equation}
\label{tN}
\varphi_{t_{N}}=e^{-\frac{i}{\hbar}{\widehat h}(t_N)\Delta t}e^{-\frac{i}{\hbar}{\widehat h}(t_{N-1})\Delta t}... e^{-\frac{i}{\hbar}{\widehat h}(t_1)\Delta t}\varphi_{t_{0}}.
\end{equation}
For the state conditioned to stay on $M^{\sigma}_{3}$, the points $\varphi_{t_{0}}, \varphi_{t_{1}}, . . . $ belong to $M^{\sigma}_{3}$ and the steps can be identified with translations in the classical space. In other words, ${\widehat h}(t_{k})={\bf \xi}_{k}{\widehat {\bf p}}$, where ${\widehat {\bf p}}$ is the momentum operator and ${\bf \xi}_{k}$ is a vector in $\R^3$. The equation (\ref{tN}) yields then the following expression:
\begin{equation}
\label{tN1}
\varphi_{t_{N}}({\bf x})=\varphi_{t_{0}}(x-{\bf \xi}_{1}\Delta t-{\bf \xi}_{2}\Delta t- ... -{\bf \xi}_{N}\Delta t).
\end{equation}
That is, the initial state is simply translated by the vector 
\begin{equation}
\label{walk}
{\bf d}_N=\sum^{N}_{k=1}{\bf \xi}_{k}\Delta t
\end{equation}
in $\R^3$.
Now, the probability distribution of steps $-\frac{i}{\hbar} {\widehat h}(t_{k+1})\varphi_{t_{k}}$ must be the conditional probability distribution under the condition that the steps take place in $T_{\varphi_{k}}M^{\sigma}_{3}$.
Because the matrix of ${\widehat h}$ is in the Gaussian unitary ensemble, the conditional probability distribution is the usual probability distribution on the subspace $T_{\varphi_{k}}M^{\sigma}_{3}=\R^3$ and the vectors ${\bf \xi}_k$ are independent and identically normally distributed random vectors. It follows that the equation (\ref{walk}) defines a random walk with Gaussian steps on $\R^3$. This is known to approximate the Brownian motion on $\R^3$ and can be used to model the process of measurement over the period of observation.

There is also a converse result. Namely, given a random walk ${\bf d}_N$ with Gaussian steps on $\R^3$, there is a unique Gaussian unitary ensemble such that the Scr{\"o}dinger evolution with Hamiltonian represented by a random matrix in this ensemble yields the walk ${\bf d}_N$ when constrained to $M^{\sigma}_{3}$. In fact, the distribution of steps in the direction tangent to $M^{\sigma}_{3}$ defines the distribution of all entries of the random matrix in the Gaussian unitary ensemble and therefore defines the ensemble completely.

Note also that the Gaussian orthogonal ensemble used in \cite{Wigner} would not result in the Brownian motion on $M^{\sigma}_{3}$ in this way. In fact, the momentum operator in the derivation is Hermitian but not orthogonal. It is known that Hamiltonians with matrices in the Gaussian unitary ensemble are not invariant with respect to time reversal. So, the fact that the Brownian motion on the classical space submanifold was derived from Schr{\"o}dinger evolution is tied to the use of a time-irreversible Hamiltonian.

Let us see now what kind of random walk of state is obtained for the state driven by the Hamiltonian ${\widehat h}$ and not constrained to $M^{\sigma}_{3}$.
First of all, we claim that for all initial states $\{\varphi\}$ in the space of states $CP^{L_2}$,
the vector $d\varphi=-\frac{i}{\hbar}{\widehat h}\varphi dt$ with such a Hamiltonian is a normal random vector in the tangent space $T_{\{\varphi\}}CP^{L_2}$ with a spherical distribution. In particular, the probability distribution of steps of the random walk is isotropic and homogeneous. 
In fact, because for all values of $t$, the matrix of ${\widehat h}$ is in the Gaussian unitary ensemble, the probability density function $P({\widehat h})$ of ${\widehat h}$ is invariant with respect to conjugations by unitary transformations. That is,  $P(U^{-1}{\widehat h}U)=P({\widehat h})$ for all unitary transformations $U$ acting in the Hilbert space of states.
Also, for all unitary transformations $U$ that leave $\{\varphi\}$ unchanged and therefore all $U$ that act in the tangent space $T_{\{\varphi\}}CP^{H}$, we have
\begin{equation}
\label{v}
(U^{-1}{\widehat h}U \varphi, v)=({\widehat h}U \varphi, Uv)=({\widehat h}\varphi, Uv),
\end{equation}
where $v$ is a unit vector in $T_{\{\varphi\}}CP^{H}$. It follows that 
\begin{equation}
\label{vv}
P({\widehat h} \varphi, v)=P({\widehat h} \varphi, Uv),
\end{equation}
where $P$ is the probability density of the components of ${\widehat h}\varphi$ in the given directions.
By a proper choice of $U$, we can make $Uv$ to be an arbitrary unit vector in $T_{\{\varphi\}}CP^{H}$, proving the isotropy of the distribution.
Furthermore, for all unitary operators $V$ in $H$ and a unit vector $v$ in $T_{\{\varphi\}}CP^{H}$, we have
\begin{equation}
\label{vw}
P({\widehat h} \varphi, v)=P(V^{-1}{\widehat h}V \varphi, v)=P({\widehat h}V\varphi, Vv).
\end{equation}
Because $V\varphi$ is an arbitrary state and $Vv$ is in the tangent space $T_{\{V\varphi\}}CP^{H}$, we conclude with the help of (\ref{v}) that the probability density function is independent of the initial state of the system, proving the homogeneity of the distribution. The components of the vector ${\widehat h}\varphi dt$ are independent and normally distributed because the entries in the columns of the matrix of ${\widehat h}$ are independent and normally distributed. It follows that $-\frac{i}{\hbar}{\widehat h}\varphi dt$ is a normal random vector with a spherical distribution.

Because the steps of the obtained walk of state are independent and the distribution of steps is homogeneous and isotropic, the probability of reaching a certain state during the given period of observation can only depend on the Fubini-Study distance between the initial and final states. As we proved, under these conditions the normal probability distribution on $M^{\sigma}_{3}$ implies the Born rule for the probability of transition between states. So, the Schr{\"o}dinger evolution with the Hamiltonian in the Gaussian unitary ensemble yields the Brownian motion of the particle whose state is constrained to $M^{\sigma}_{3}$ and the Born rule for the probability of transition between unconstrained states of the particle in $CP^{L_{2}}$. 

This result combined with the obtained relationship between Newtonian and Schr{\"o}dinger dynamics is telling us that in the considered model there is no fundamental difference between macroscopic and microscopic particles. The dynamics of macroscopic particles, including their behavior under a measurement is identical to the corresponding dynamics of microscopic particles whose state is constrained to the classical space submanifold of the space of states. Moreover, knowing the dynamics of a macrosystem, whether the system is measured or not, we can now predict in a unique way the dynamics of the corresponding microsystem, and vice versa. 

To obtain a unified mechanism of measurement in classical and quantum mechanics, we still need to explain why in the model the state of a macroscopic particle is constrained to $M^{\sigma}_{3}$. We also need to understand the dynamics of the system consisting of  a measured particle and a measuring device during measurement. Finally, the validity of the Born rule for the system driven by the Hamiltonian in ${\bf (RM)}$ comes with the downside that numerous final states are equally likely to occur in a measurement. We then need to explain why, for instance, a measurement of the position of a particle yields only the states of a well-defined position of the particle in $\R^3$. 

%
%
To address the first question, note that the motion of state driven by the Hamiltonain ${\widehat h}(t)$ represented by a random matrix in the Gaussian unitary ensemble is essentially the Brownian motion on the space of states $CP^{L_{2}}$. For states constrained to $M^{\sigma}_{3}$, this motion reduces to the Brownian motion of the particle on $\R^3$. As is well known, the physical Brownian motion of sufficiently large particles in a medium in $\R^3$ is trivial. This is because the total force acting on such particles from atoms and molecules of the medium vanishes. In this case, the diffusion coefficient for the process is zero and the particle remains at rest in the lab system. Therefore, the action of the Hamiltonian ${\widehat h}(t)$ in {\bf (RM)} that represents this situation in the direction of vectors in the tangent space $T_{\{\varphi\}}M^{\sigma}_{3}$ at initial point $\{\varphi\}$ in $M^{\sigma}_{3}$ must be nearly trivial. However, because the distribution of steps for the Hamiltonian in the Gaussian unitary ensemble is isotropic and homogeneous, the distribution of all entries of ${\widehat h}(t)$ must be centered at zero and have a vanishingly small variance.  In other words, if under these conditions the particle does not move in $\R^3$, then the state of the particle does not move in the direction of vectors in $T_{\{\varphi\}}M^{\sigma}_{3}$, and then it cannot move in any other direction in the tangent space to the space of states at $\{\varphi\}$. This reflects the fact that the state under these conditions is pushed simultaneously in all possible directions in the space of states, so that the net displacement of state is zero in the lab system.  


If an external potential $V({\bf x})$ is applied to such a particle, the state of the particle will evolve in the classical phase space submanifold $M^{\sigma}_{3,3}$ in accord with Newtonian dynamics. 
%
%
This can be shown by identifying components of the velocity of state $d\varphi/dt$ under the Hamiltonian ${\widehat h}+V$, as in \cite{KryukovPosition}. The underlying assumption is that the operator ${\widehat h}$ is built in the usual way from the kinetic and potential energy terms of all participating particles.
The effect of the external potential on the velocity appears in the form of components tangent to the classical phase space submanifold $M^{\sigma}_{3,3}$. The state is pushed along the submanifold and the particle moves classically in $\R^3$. Alternatively, note that in a frame moving relative to the lab system  with linear acceleration ${\bf w}$, the Schr{\"o}dinger equation acquires an extra term $V=-m{\bf w}\cdot {\bf x}$ \cite{Mashhoon, Green}, observed in the experiment \cite{Wrob}. However, any differentiable potential is approximately linear within small distances. So, the action of an external potential on states in $M^{\sigma}_{3,3}$ with small $\sigma$ is equivalent to the transformation to an accelerated reference frame, the change that preserves $M^{\sigma}_{3,3}$. It follows that the state initially in $M^{\sigma}_{3,3}$ remains confined to $M^{\sigma}_{3,3}$ while the particle moves classically with acceleration ${\bf w}=-\nabla V/m$ in $\R^3$.
This explains why the state of a macroscopic measuring device positioned initially in the classical phase space submanifold of the space of states remains confined to the submanifold and obeys the laws of Newtonian dynamics. 

Furthermore, consider a microscopic particle in a medium, such as gas or liquid. 
A small neutral particle in the medium may be able to move freely through the medium.
On another hand, the Brownian-size particles would experience a Brownian motion in the medium. Per {\bf (RM)}, the smaller microscopic particles whose interaction with atoms and molecules of the medium cannot be neglected, will evolve by the Hamiltonian represented by a random matrix. The state of a particle will perform a random walk in $CP^{L_{2}}$. For the steps within the submanifold $M^{\sigma}_{3}$, the walk represents the Brownian motion on $\R^3$.

Suppose we increase the size of the particle. The Brownian motion will eventually stop, which also means that the state of the particle in $CP^{L_{2}}$ in the model will remain fixed in $M^{\sigma}_{3,3}$. The size of the particle needed for this to happen depends on physical properties of the medium such as its dynamic viscosity and temperature. In particular, the boundary between microscopic and macroscopic bodies in the model depends on the medium the bodies are in. The cosmic background radiation is probably the ultimate medium to define such a boundary and to test for it. The state of a particle that is macroscopic for a given medium is confined to  the submanifold $M^{\sigma}_{3,3}$ and moves in accord with Newtonian dynamics. We interpret such a motion as the classical motion of particle in $\R^3$. 
Under proper conditions, the macroscopic particle may again experience a Brownian-like motion in $\R^3$, this time purely classical. A large ball kicked by many feet from all sides would be a large-scale example of such a motion. In this case, the state of the particle is already constrained to $M^{\sigma}_{3,3}$ and its evolution is driven by the classical Hamiltonian. 

Consider now a system consisting of a light particle interacting with a much heavier particle. In Newtonian physics, the effect of the light particle on the motion of the heavy particle can be neglected. In particular, the Brownian motion of the heavy particle due to its interaction with the surroundings will not be altered by its interaction with the light particle. 
From {\bf (RM)}, it follows that the state of the heavier particle for the corresponding system of two microparticles in the model exists as a random variable with values in the space of states and is driven by a random Hamiltonian.
In particular, the state of the system of two particles must be the product of states of the particles, i.e., the state will remain separable throughout the evolution.
We know that when the heavier particle is sufficiently large, its state will be confined to the submanifold $M^{\sigma}_{3,3}$ and the motion of the particle will be classical. 

In particular,  the system consisting of a microscopic particle and a macroscopic measuring device in the model will always be in a product state with the state of the device constrained to the submanifold $M^{\sigma}_{3,3}$ and satisfying the classical Newtonian dynamics. The microscopic particle in this case will move in accord with the Schr{\"o}dinger dynamics in the surroundings provided by the device and the environment. When the measurement occurs, the state of the particle experiences a random walk that satisfies the Schr{\"o}dinger equation with Hamiltonian in {\bf (RM)}. The walk results in the Born rule for the probability of transition to the observed state.





It remains to understand why in a measurement described by the model we can only observe the eigentstates of a measured quantity. Indeed, this seems to contradict the result that the state driven by the Hamiltonian in {\bf (RM)} is equally likely to reach any state at a given Fubini-Study distance from the initial state. To answer, consider first the classical experiment of firing bullets into a target and measuring position of the bullets at rest. For the bullet to hit the target, the experimenter must properly aim the gun. Without this, the probability of hitting the target is small. 
We claim that the role of the measuring device in quantum mechanics in the proposed model is similar: it makes the state approach the target and records the outcome. The difference is that now the process takes place in the space of states.

Consider for example a setup where the position of an electron is measured by a scintillating screen, as in the double-slit experiment. What does it mean for the electron's state to approach the screen? Suppose the screen occupies region 
$D$ in $\R^3$.  Then position of the particles of the screen is well defined and the state of each particle can be identified with a point $g_{{\bf a},\sigma}$ in  $M^{\sigma}_{3}$, with ${\bf a}$ in $D$. The state $\Psi$ of the screen is the product of states $g_{{\bf a},\sigma}$, one for each of the particles of the screen. It belongs to the submanifold $M^{\sigma}_{3N}=M^{\sigma}_{3} \otimes ... \otimes M^{\sigma}_{3}$ in the $N$-particle space of states.  
Because the screen is macroscopic, as has been seen previously, the state of the particle-screen system at any time  is a product $\varphi \otimes \Psi$. After the measurement, the state of the system is $g_{{\bf c},\sigma} \otimes \Psi$, where $g_{{\bf c},\sigma}$ is the state of the measured particle and the internal changes in the screen are not considered. It follows that the state  $\varphi \otimes \Psi$ is close to the end-state  $g_{{\bf c},\sigma} \otimes \Psi$ exactly when the state $\varphi$ is close to $g_{{\bf c},\sigma}$ in the Fubini-Study metric. So, in the process of measurement, the state $\varphi$ of the particle must approach the classical space submanifold $M^{\sigma}_{3}$ in $CP^{L_{2}}$ and the subset of states $g_{{\bf a},\sigma}$ with ${\bf a}$ in $D$ in it. 

The state $g_{{\bf c},\sigma}$ with a proper value of $\sigma$ can be identified with the lowest energy state of the electron in the potential near the point of the screen where the electron is absorbed. The initial state $\varphi$ can be decomposed into a superposition of energy eigenstates in this potential. If the coefficients of the higher-energy components of $\varphi$ are not small, the state is away from $M^{\sigma}_{3}$ in the Fubini-Study metric. To approach $M^{\sigma}_{3}$, the electron must loose energy so that the higher-energy components must become small or vanish completely. The process is similar to the one with the bullet hitting the target.  
This time, the stopping power of the screen is responsible for lowering the energy of the electron. Aiming, or providing the electron with a proper momentum directed towards the screen in $\R^3$ is also needed to ensure that the electron lands on the screen. However, while the motion of the bullet's state is happening in $M^{\sigma}_{3}$ and is identified with a motion in $\R^3$, the motion of the electron's state is happening in $CP^{L_{2}}$ at large. The process can be described by the Schr{\"o}dinger equation with elements of the quantum field theory to account for the interaction of the electron with the electromagnetic field and for the indistinguishability of electrons. When the electron is absorbed, only the lowest energy component $g_{{\bf c},\sigma}$ survives, identifying the position of the electron on the screen.

%

Mathematically, modeling the motion of state towards the classical space $M^{\sigma}_{3}$ and the screen 
amounts to adding a drift term to the random walk specified in {\bf (RM)}. 
The drift ensures that the state arrives to the screen, while the random walk delivers the Born rule for the possible outcomes $g_{{\bf c},\sigma}$. A clever design of the measuring device ensures a proper drift, so that the state is guaranteed to arrive to the needed part of the space of states. As a result, the probability to find the electron at a certain point of the screen is the conditional probability under the condition that the electron hit the screen. This explains the issue of observability of just a limited subset of all possible states in a measurement in the model.

Let us summarize the lessons learnt from the proposed model of measurement. First, to answer the question in the title of the paper, the Schr{\"o}dinger dynamics {\it is} capable of explaining the process of measurement. This is achieved by replacing the deterministic evolution of state with the motion driven by the Hamiltonian satisfying {\bf (RM)}. In this case the distribution of end-states satisfies the Born rule for all initial states. The linear property of the evolution does not cause a problem because the distribution is homogeneous and isotropic. In particular, the random walk of a linear superposition of states satisfies the same Born rule. 
Second, the Newtonian dynamics of macroscopic bodies is identified with the Schr{\"o}dinger dynamics constrained to the classical phase space submanifold 
in the space of states. 
Likewise, the motion of a Brownian particle in $\R^3$ follows from the Schr{\"o}dinger evolution with Hamiltonian in {\bf (RM)}, constrained to the classical space submanifold $M^{\sigma}_{3}$. This also points to the origin of the constraint: when the particle is sufficiently large so that its Brownian motion in the given medium trivializes, the state of the particle ``freezes up" on $M^{\sigma}_{3}$, leading to its classical behavior in an external potential.
Third, the irreversibility of classical and quantum measurements in the model is tied to time-irreversibility of the Hamiltonian in {\bf (RM)}. Namely, the derivation of the Brownian motion from the Schr{\"o}dinger evolution requires that the matrix that represents the Hamiltonian is in the Gaussian unitary ensemble. Such Hamiltonians do not commute with the time reversal operator, potentially providing the origin of the irreversibility.
Fourth, under {\bf (RM)}, the state of the system consisting of a microscopic and a macroscopic particle remains separable throughout the evolution. The macroscopic measuring device behaves classically during the measurement, while the state of the microscopic particle experiences a random walk leading to the Born rule for the probability of transition to the end-states. 
Finally, the role of the measuring device in the model consists in creating a drift of state towards the device and in recording the result of observation.

The crux of the obtained results is the extension of the motion of particles in the classical space to the motion of states in the space of states. Going from the motion in the classical space submanifold of the space of states to the motion in the full space of states is what allowed us to connect the classical and the quantum in such a beautiful and concise way. Based on that, we  hypothesize that the space of states rather than the classical space is the true arena for all physical processes.

\end{document}